\documentclass[12pt]{article}
\usepackage{amssymb,amsmath}
\usepackage[noblocks]{authblk}
\usepackage[top=0.75in, bottom=0.75in, left=0.75in, right=0.75in, dvips]{geometry}
\usepackage{caption}
\usepackage{graphicx}
\begin{document}
\textwidth 10.0in 
\textheight 9.0in 
\topmargin -0.60in
\title{Renormalization Scheme Dependence and Renormalization Group Summation}
\author[1]{F.A. Chishtie}
\author[2,3]{D.G.C. McKeon}
\affil[1] {Department of Physics and Astronomy, The
University of Western Ontario, London, ON N6A 3K7, Canada} 
\affil[2] {Department of Applied Mathematics, The
University of Western Ontario, London, ON N6A 5B7, Canada} 
\affil[3] {Department of Mathematics and
Computer Science, Algoma University, \newline Sault St.Marie, ON P6A
2G4, Canada}
\date{}
\maketitle    

\maketitle
\noindent
email: dgmckeo2@uwo.ca

\begin{abstract}
We consider logarithmic contributions to the free energy, instanton effective action and Laplace sum rules in QCD that are a consequence of radiative corrections. Upon summing these contributions by using the renormalization group, all dependence on the renormalization scale parameter $\mu$ cancels. The renormalization scheme dependence in these processes is examined, and a renormalization scheme is found in which the effect of higher order radiative corrections is absorbed by the behaviour of the running coupling.
\end{abstract}

\section{Introduction}
The process of renormalization in quantum field theory induces at finite order in perturbation theory a dependence on arbitrary parameters.  The requirement that physical processes be independent of these parameters leads to the renormalization group (RG) equations [1-3].  Irrespective of the renormalization scheme (RS) used to excise divergences from a calculation, one of these arbitrary parameters is the renormalization mass scale parameter $\mu$.  When one uses a mass-independent RS [4-5] there is also arbitrariness associated with the expansion coefficients of the RG functions associated with the couplings [6] and mass parameters [7] occurring in the theory. The RS dependence of the cross section for $e^+e^-$ scattering [8,9] as well as the semi-leptonic decay rate of the $b$ quark [10] has been considered.

In this paper we apply the approach used in refs. [8-10] to a number of other processes. In particular, we consider the free energy in thermal QCD, the effective action induced by instantons, and the Laplace sum rules for scalar gluon currents in QCD.  The point of the calculations is to illustrate two things. The first is to show that if the RG equation associated with the arbitrary mass scale $\mu$ can be used to sum all logarithmic corrections to a processes that are due to radiative effects, then the explicit and implicit dependence on $\mu$ entirely cancels.  The second is to show that the RS dependency of these RG summed results can be exploited to reduce the perturbative contributions to a finite number of terms in a series expansion in powers  of the coupling; all higher loop calculations then serve to determine the behaviour of the running parameters that characterize the theory.

\section{Thermal Free Energy}
The thermal free energy in QCD with $n_f$ quark flavours at temperature $T$ is given by [11-13],
\begin{equation}
\mathcal{F} = \mathcal{F}_0 \left[1 + \sum_{n=0}^\infty \left( R_n (U) a^{n+1} +  S_n(U) a^{n + 3/2} +  T_n(U)a^{n+2}\ln a \right)\right]
\end{equation}
with 
\begin{equation}
R_n(U) =  \sum_{m=0}^\infty r_{n+m,m} U^m, \quad S_n(U) =  \sum_{m=0}^\infty s_{n+m,m} U^m, \quad T_n(U) =  \sum_{m=0}^\infty t_{n+m,m}U^m \nonumber
\end{equation}
where $a = \alpha_s/\pi$ is the couplant,  $\mathcal{F}_0$ is the ideal gas value and $U = a \ln \left(\frac{\mu}{2\pi T}\right) \equiv a L$ (one could always absorb the factor of $\ln(2\pi)$ appearing in $U$ into a redefinition of the coefficients $r_{mn}$, $s_{mn}$ and $t_{mn}$). At finite order, this result shows strong dependence on the renormalization scale $\mu$, both through the explicit dependence on  $\ln \frac{\mu}{2\pi T}$, and implicitly through $a(\mu)$ where
\begin{align}
\mu \frac{da}{d\mu} &= \beta(a)\nonumber\\
&= -ba^2 (1 + ca + c_2a^2 + \ldots).
\end{align}
$\mathcal{F}$ has been computed when the sums in eq. 
(1) extend just to $1$. The exact expression for $\mathcal{F}$ is independent of $\mu$ and so we have the RG equation
\begin{equation}
\left( \mu \frac{\partial}{\partial\mu} + \beta(a) \frac{\partial}{\partial a}\right) \mathcal{F} = 0.
\end{equation}

In ref. [14] it was shown that eq. (3) allows one to compute $R_0$, $S_0$, $T_0$ (the ``leading-log'' - $LL$) result once $b$ is known; from this and a knowledge of $c$ then $R_1$, $S_1$, $T_1$ (the ``next-to-leading-log'' - $NLL$) are determined.  Following this, then $R_n$, $S_n$, $T_n$ (the $N^nLL$ result) are fixed by a set of nested equations provided $c_n$ is known (with the computed values of $R_n(0)$, $S_n(0)$, $T_n(0)$ serving as boundary conditions).  These RG summed results, being closer to the exact value of $\mathcal{F}$ than purely perturbative results, exhibit a diminished dependence on $\mu$.

We now reconsider $RG$ summation of $\mathcal{F}$, following the approach of refs. [8,9].  If we now rewrite eq. (1) in the form
\begin{equation}
\mathcal{F} = \mathcal{F}_0  \left( 1 + \sum_{n=0}^\infty  \Omega_n (a) L^n\right)
\end{equation}
where 
\begin{equation}
\Omega_n(a) =  \displaystyle{\sum_{m=0}^\infty} \left[ r_{n+m,n}a^{n+m+1} + s_{n+m,n}a ^{n+m+\frac{3}{2}} + t_{n+m,n}a^{n+m+2} \ln a\right]\nonumber
\end{equation}
then together eqs. (3-4) result in
\begin{equation}
\Omega_n (a) = - \frac{\beta}{n}\frac{d}{da}\Omega_{n-1}(a),
\end{equation}
which by eq. (2) can be written as
\begin{equation}
\Omega_n \left(a \left( \ln \frac{\mu}{\Lambda}\right)\right) = -\frac{1}{n} \frac{d}{d\ln\left(\frac{\mu}{\Lambda}\right)} \Omega_{n-1} 
\left(a \left( \ln \frac{\mu}{\Lambda}\right)\right)
\end{equation}
where $\Lambda$ is related to be boundary condition on eq. (2) [6],
\begin{equation}
\ln \left( \frac{\mu}{\Lambda}\right) = \int_0^a \frac{dx}{\beta (x)} + \int_0^\infty \frac{dx}{bx^2(1+cx)}.
\end{equation}
and where $b$ and $c$ appear in eq. (2).  Together, eqs. (4,6) result in
\begin{align}
\mathcal{F}/\mathcal{F}_0  &= \left[ 1 + \sum_{n=0}^\infty \frac{(-L)^n}{n!}
 \left( \frac{d}{d\left(\ln \frac{\mu}{\Lambda}\right)}\right)^n \Omega_0
  \left(a  \left( \ln \frac{\mu}{\Lambda}\right)\right)\right]\\
  &= \left[ 1 + \Omega_0 \left( a \left(\ln \frac{\mu}{\Lambda} - \ln \frac{\mu}{2\pi T}\right)\right)\right]\nonumber \\
  &= \left[ 1 + \Omega_0 \left( a \left(\ln \frac{2\pi T}{\Lambda}\right)\right)\right].
\end{align}
In eq. (9), all dependence of $\mathcal{F}$ on $\mu$ (both implicit and explicit) has cancelled.

In Fig. 1, a graph of $\mathcal{F}/\mathcal{F}_0 $ vs $T$ is shown with $\Lambda = 300 MeV$ for the RG summed and scale-independent result of eq. (9), the $LL$ and $NLL$ RG summed result with $\mu=1 GeV$ and $\mu=4 GeV$ as well as the perturbative results of eq. (1) with $n=0, 1$. This shows that the pertubative results along with the RG summed are fairly dependent on the RG scale $\mu$, whereas the scale independent eq. (9) (which is in the MS scheme) demonstrates certain regions of T where it is bounded by the RG summed results, while being different, especially for low values of $T$. 

Together eqs. (1,4) show that eq. (9) is of the form 
\begin{equation}
\mathcal{F} = \mathcal{F}_0  \left[ 1 + \sum_{n=0}^\infty \left( A_n a^{n+1} + B_n a^{n+3/2} + C_n a^{n+2} \ln a\right)\right]
\end{equation}
where $A_n = R_n(0)$, $B_n = S_n(0)$, $C_n = T_n(0)$ and $a = a\left( \frac{2\pi T}{\Lambda}\right)$. 

The free energy $\mathcal{F}$ is also independent of the $RS$ used.  In any mass-independent renormalization scheme, $b$ and $c$ in eq. (2) are $RS$ independent while the expansion parameters $c_i(i \geq 2)$ can be used to characterize the $RS$ used. If
\begin{equation}
\frac{da}{dc_i} = \sigma_i(a) \qquad (i \geq 2)
\end{equation} 
then from the requirement
\begin{equation}
\left( \frac{\partial}{\partial\mu} \frac{\partial}{\partial c_i} -
\frac{\partial}{\partial c_i} \frac{\partial}{\partial\mu} \right) a = 0
\end{equation}
it follows that [6]
\begin{align}
\sigma_i(a) &= -b \beta(a) \int_0^a dx \frac{x^{i+2}}{\beta^2(x)}\\
&\approx a^{i+1} \left[ \frac{1}{i-1}-c \left(\frac{i-2}{i(i-1)}\right) a + \frac{1}{i+1} \left( c^2 \frac{i-2}{i} - c_2 \frac{i-3}{i-1}\right) a^2 + \ldots \right].
\end{align}
We now can use eqs. (10,13) to see how $A_n$, $B_n$ $C_n$ all depend on $c_i$. We see that if
\begin{align}
\frac{d\mathcal{F}}{dc_i} = 0 &= \mathcal{F}_0 \sum_{n=0}^\infty \Bigg\{ \frac{\partial A_n}{\partial c_i} a^{n+1} + \frac{\partial B_n}{\partial c_i} a^{n+\frac{3}{2}} + 
\frac{\partial C_n}{\partial c_i} a^{n+2} \ln a  \\
&+ \sigma_i(a) \Bigg[ (n+1) A_n a^n + \left( n + \frac{3}{2}\right) B_n a^{n + \frac{1}{2}}  \nonumber \\
 &\qquad \qquad +  \left((n+2) \ln a + 1 \right) a^{n+1} C_n \Bigg] \Bigg\} \nonumber
\end{align}
is satisfied at each order in $a^m \ln^n a$, it follows that
\begin{equation}\tag{16a-c}
\frac{\partial A_0}{\partial c_i} = \frac{\partial B_0}{\partial c_i} = \frac{\partial C_0}{\partial c_i} = 0
\end{equation}
\begin{equation}\tag{17a-c}
\frac{\partial A_1}{\partial c_i} = \frac{\partial B_1}{\partial c_i} = \frac{\partial C_1}{\partial c_i} = 0
\end{equation}
\begin{equation}\tag{18a-c}
\frac{\partial A_2}{\partial c_i}  + A_0\delta_2^i =  \frac{\partial B_2}{\partial c_i} + \frac{3}{2}  B_0\delta_2^i = \frac{\partial C_2}{\partial c_i} +  2C_0\delta_2^i= 0
\end{equation}
\begin{align}\tag{19a-c}
\frac{\partial A_3}{\partial c_i} & + (2A_1 + C_0)\delta_2^i + \frac{1}{2} A_0\delta_3^i = 
 \frac{\partial B_3}{\partial c_i} + \frac{5}{2}  B_1\delta_2^i +  \frac{3}{4}  B_0\delta_3^i \\
& = \frac{\partial C_3}{\partial c_i} +  3C_1\delta_2^i + C_0\delta_3^i = 0\nonumber
\end{align}
etc.\\
Eqs. (16-19) can be integrated to give
\begin{equation}\tag{20a-c}
A_{0,1} = \alpha_{0,1}\qquad B_{0,1} = \beta_{0,1} \qquad C_{0,1} = \gamma_{0,1}
\end{equation}
\begin{equation}\tag{21a-c}
A_2 = \alpha_2 - \alpha_0 c_2 \qquad    B_2 = \beta_2 - \frac{3}{2}\beta_0 c_2 \qquad 
C_2 = \gamma_2 - 2\gamma_0 c_2  
\end{equation}
\begin{equation}\tag{22a-c}
A_3 = \alpha_3 - (2\alpha_1 +\gamma_0)c_2 - \frac{1}{2} \alpha_0 c_3 \quad
B_3 = \beta_3 - \frac{5}{2} \beta_1 c_2 - \frac{3}{4} \beta_0 c_3 \quad
C_3 = \gamma_3 - 3\gamma_1 c_2 - \gamma_0 c_3
\end{equation}
etc.\\
In eqs. (20-22), $\alpha_k$, $\beta_k$ and $\gamma_k$ are constants of integration and consequently are $RS$ invariants.  Once $A_k$, $B_k$, $C_k$ and $\beta(a)$ have been explicitly calculated in a particular $RS$ to order $N$ in the loop expansions, then $\alpha_k$, $\beta_k$, $\gamma_k$ can be found from eqs. (20-22) to order $N-1$.  Two particular schemes then suggest themselves. One is the `` 't Hooft scheme'' [15] in which $c_k = 0 (k \geq 2)$ in which case $A_k = \alpha_k$, $B_k = \beta_k$ and $C_k = \gamma_k$. A second scheme is one in which either $A_k$, $B_k$ or $C_k$ vanish for $k \geq 2$.  If $A_k = 0 (k \geq 2)$ then by eqs. (21,22) the $\beta$ function expansion coefficients in eq. (2) are fixed,
\begin{equation}\tag{23a,b}
c_2 = \frac{\alpha_2}{\alpha_0}\qquad c_3 = \frac{2}{\alpha_0} \left[\alpha_3 - (2\alpha_1 + \gamma_0) \left( \frac{\alpha_2}{\alpha_0}\right) \right]
\end{equation}
etc.\\
and so
\begin{equation}\tag{24a,b,c}
A_2 = 0, \quad B_2 = \beta_2 - \frac{3}{2} \frac{\beta_0\alpha_2}{\alpha_0}, \quad C_2 = \gamma_2 - 2 \frac{\gamma_0\alpha_2}{\alpha_0}
\end{equation}
etc. A similar approach can be used when considering a $RS$ in which $B_k = 0 (k \geq 2)$ or $C_k = 0 (k \geq 2)$.

\section{Instanton Effective Action}

The form of the effective action in an $SU(2)$ gauge theory with $n_f$ flavours is
\begin{equation}\tag{25}
\mathcal{L}_{eff} \sim K \int_0^\infty d\rho \; \rho^{-5+3n_f} \exp \left\lbrace -\frac{8\pi^2}{g^2} S \right\rbrace \left( a \equiv \frac{g^2}{4\pi^2}\right).
\end{equation}
In eq. (25), the scale parameter $\rho$ is the instanton size and $S$ has been computed to one loop order [16].  In general, $S$ has the form
\begin{equation}\tag{26}
S = \sum_{n=0}^\infty \sum_{m=0}^n T_{n,m} a^n \ln^m (\mu\rho) \quad (T_{0,0} = 1).
\end{equation}
Since $\mathcal{L}_{eff}$ satisfies the $RG$ equation
\begin{equation}\tag{27}
\left( \mu \frac{\partial}{\partial \mu} + \beta(a) \frac{\partial}{\partial a}\right)\mathcal{L}_{eff} = 0
\end{equation}
the functions $S_n(aL)\;\;(L \equiv \ln (\mu\rho))$ can be calculated iteratively where
\begin{equation}\tag{28}
S_n(aL) = \sum_{m=0}^\infty T_{n+m,m}(aL)^m \quad (S_n(0) = T_{n,0}).
\end{equation}
This has been considered in ref. [17]. However, if only a finite number of terms are kept in the expansion of eq. (26), $S$ will retain at least a residual dependence on $\mu$.

This dependence can be shown to cancel if one were to write (in much the same way as eq. (4)),
\begin{equation}\tag{29}
S = \sum_{n=0}^\infty A_n(a) L^n
\end{equation}
where now
\begin{equation}\tag{30}
A_n(a) = \sum_{m=0}^\infty T_{n+m,n} a^{n+m} \quad (n = 0,1,2,\ldots)\;.
\end{equation}
Then as by eq. (27)
\begin{equation}\tag{31}
\left( \mu \frac{\partial}{\partial \mu}+ \beta(a) \frac{\partial}{\partial a}\right) \left( \frac{1}{a} \sum_{n=0}^\infty A_n (a) L^n\right) = 0
\end{equation}
it follows that
\begin{equation}\tag{32}
\frac{1}{a} A_n(a) = \frac{-1}{n} \beta(a) \frac{d}{da}\left(\frac{1}{a} A_{n-1} (a)\right)
\end{equation}
and so, just as eq. (9) follows from eq. (5), we see that eq. (32) results in
\begin{equation}\tag{33}
\mathcal{L}_{eff} = K \int_0^\infty dp \;\rho^{-5+3n_f} \exp \left\lbrace
\frac{-2}{a(\ln\frac{1}{\rho\Lambda})} A_0 \left( a\left(\ln \frac{1}{\rho\Lambda}\right)\right)\right\rbrace\;.
\end{equation}
As with eq. (9), all the implicit and explicit dependence of $\mathcal{L}_{eff}$ on $\mu$ has cancelled upon doing the $RG$ sum.

We now see that by eq. (26)
\begin{equation}\tag{34}
\frac{1}{a} A_0 (a) = \frac{1}{a}\left( 1 + \sum_{n=1}^\infty T_n a^n\right)
\end{equation}
where  $a = a\left(\ln \frac{1}{\rho\Lambda}\right)$ and $T_n \equiv T_{n,0}$.  As $\mathcal{L}_{eff}$ is $RS$ independent, then as with eq. (15)
\begin{equation}\tag{35}
\frac{d}{dc_i} \left(\frac{1}{a} A_0 (a) \right) = 0 = \sum_{n=1}^\infty \left( \frac{\partial T_n}{\partial c_i} a^{n-1}\right) + \sigma_i(a) \left( \frac{-1}{a^2} + \sum_{n=1}^\infty (n-1) T_n a^{n-2} \right)
\end{equation}
which is satisfied order by order in $a$ provided
\begin{equation}\tag{36a}
\frac{\partial T_1}{\partial c_i} = 0
\end{equation}
\begin{equation}\tag{36b}
\frac{\partial T_2}{\partial c_i} - \delta_2^i = 0
\end{equation}
\begin{equation}\tag{36c}
\frac{\partial T_3}{\partial c_i} - \frac{1}{2}\delta_3^i = 0
\end{equation}
\begin{equation}\tag{36d}
\frac{\partial T_4}{\partial c_i} - \frac{1}{3}\delta_4^i + \frac{c}{6} \delta_3^i + \left( T_2 - \frac{c_2}{3}\right) \delta_2^i = 0
\end{equation}
etc.\\
Solving for $T_i$ in turn from eq. (36) leads to
\begin{equation}\tag{37a}
T_1 = \tau_1
\end{equation}
\begin{equation}\tag{37b}
T_2 = \tau_2 + c_2
\end{equation}
\begin{equation}\tag{37c}
T_3 = \tau_3 + \frac{c_3}{2}
\end{equation}
\begin{equation}\tag{37d}
T_4 = \tau_4 + \frac{c_4}{3} - \frac{cc_3}{6} - \frac{c_2^2}{3} - \tau_2 c_2
\end{equation}
etc.\\
In eq. (37), the $\tau_i$ are constants of integration and hence are $RS$ invariants. The 't Hooft $RS$ in which we choose $c_i = 0 (i \geq 2)$ leads to 
\begin{equation}\tag{38}
T_i = \tau_i \quad (i \geq 1).
\end{equation}
In a second $RS$, $T_i = 0\;\; (i \geq 2)$ so that
\begin{equation}\tag{39a}
c_2 = -\tau_2
\end{equation}
\begin{equation}\tag{39b}
c_3 = -2\tau_3
\end{equation}
\begin{equation}\tag{39c}
c_4 = -3 \left( \tau_4 + \frac{c\tau_3}{3} + \frac{2\tau_2^2}{3}\right)
\end{equation}
etc.\\
with
\begin{equation}\tag{40}
S = 1 + \tau_1 a\left( \ln \frac{1}{\rho\Lambda}\right)
\end{equation}
there being no higher order contributions to $S$ in this scheme.

Substitution of eq. (40) into eq. (25) leads to 
\begin{equation}\tag{41}
\mathcal{L}_{eff} \sim K \int_0^\infty d\rho \;\rho^{-5+3n_f} \exp - 2 \left[\frac{1}{a(\ln \frac{1}{\rho\Lambda})} + \tau_1 \right].
\end{equation}
As $\rho \rightarrow 0^+$, at one loop order by eqs. (2,7)
\begin{equation}\tag{42}
a\left( \ln \frac{1}{\rho\Lambda} \right) \longrightarrow 
\frac{1}{b\ln (\frac{1}{\rho\Lambda})}
\end{equation}
on account of asymptotic freedom, and thus at the lower end of integration in eq. (41)
\begin{equation}\tag{43}
\rho^{-5+3n_f} \exp\left( \frac{-2}{a(\ln \frac{1}{\rho\Lambda})} \right) \longrightarrow \rho^{-5+3n_f} (\rho\Lambda)^{2b}.  
\end{equation}
As for $SU(2)$, $b = \left(\frac{22-2n_f}{6} \right)$ we see that the integral in eq. (41) is convergent at the lower limit of integration for all values of $n_f$ in this $RS$.  As $\rho \rightarrow \infty$, we anticipate that $1/a \rightarrow 0$ and so for $n_f$ sufficiently small we see that at the upper limit of integration the integral in eq. (41) also converges.  Having the integral in eq. (41) converge at both limits of integration is not what one expects when considering one-loop contributions to $S$ [16] or the leading-log contributions to $S$ [17].

\section{QCD Laplace Sum Rules for Scalar Gluon Currents}

With the scalar gluon operator
\begin{equation}\tag{44}
j_G(x) = - \frac{2\pi\beta(a)}{ab} G_{\mu\nu}^a (x) G^{a\mu\nu} (x)
\end{equation}
the correlation function
\begin{equation}\tag{45}
\Pi_G (p^2) = i \int dy\; e^{ip\cdot y} < 0| T j_G(y)j_G(0)|0>
\end{equation}
is used to define the Laplace sum rule
\begin{equation}\tag{46}
\mathcal{L}_k^{pert}(\tau) = \frac{1}{\pi} \int_0^\infty ds\; s^{k+2} e^{s\tau} I_M\Pi_G^{pert} (s).
\end{equation}
It is now possible to make the expansion [18,19]
\begin{equation}\tag{47}
\mathcal{L}_k^{pert}(\tau) =  \frac{\mbox{\Large{\textit{a}}}_k a^2}{\tau^{k+3}} \left[ 1 + \sum_{n=1}^\infty \sum_{m=0}^n T_{n,m}^{(k)} a^n L^m\right] \quad (L = \ln(\sqrt{\tau} \mu)). 
\end{equation}
As with $\mathcal{F}$ in eq. (1) and $\mathcal{L}_{eff}$ (25), $\mathcal{L}_k^{pert}$ in eq. (46) has explicit and implicit dependence on $\mu$ at any finite order of perturbation theory that cancels upon summing all perturbative effects.  In ref. [19] it is shown how summation of $N^pLL$ contributions to $\mathcal{L}_k^{pert}$ (using the computed values of $b,c,\ldots c_p$ in eq. (2) and $T_{p,0}^{(k)}$ in eq. (47)) considerably reduces the $\mu$ dependence of $\mathcal{L}_k^{pert}$ from that of the purely perturbative result to order $(p+1)$ in the loop expansion.  It is now possible to show how all of the $\mu$ dependence of $\mathcal{L}_k^{pert}$ can be made to cancel upon summing all logarithmic effects.  To do this, we follow eqs. (4,29) and define 
\begin{equation}\tag{48}
A_n^{(k)} (a) = \sum_{m=0}^\infty T_{n+m,n}^{(k)} a^{n+m+2} \quad (n = 0,1,2, \ldots)
\end{equation}
where $T_{0,0} = 1$ so that eq. (47) becomes
\begin{equation}\tag{49}
\mathcal{L}_k^{pert} (\tau) = \frac{\mbox{\Large{\textit{a}}}_k}{\tau^{k+3}} \sum_{n=0}^\infty A_n^{(k)} (a) L^n.
\end{equation}
The $RG$ equation
\begin{equation}\tag{50}
\left( \mu \frac{\partial}{\partial \mu} + \beta(a) \frac{\partial}{\partial a}\right)
\mathcal{L}_k^{pert} (\tau) = 0
\end{equation}
leads to (as with eq. (6))
\begin{equation}\tag{51}
A_n^{(k)} \left( a \left( \ln \frac{\mu}{\Lambda}\right)\right) =\frac{-\beta(a)}{n} \frac{d}{a(\ln\frac{\mu}{\Lambda})} A_{n-1}^{(k)} \left(a\left(\ln\frac{\mu}{\Lambda}\right)\right)
\end{equation}
and so the sum in eq. (49) results in
\begin{equation}\tag{52}
\mathcal{L}_k^{pert} (\tau) = \frac{\mbox{\Large{\textit{a}}}_k}{\tau^{k+3}} A_0^{(k)} \left(a\left(\ln\frac{1}{\sqrt{\tau}\Lambda}\right)\right)
\end{equation}
much like eqs. (9,33) above.  By eq. (48), eq. (52) becomes
\begin{equation}\tag{53}
\mathcal{L}_k^{pert} (\tau) = \frac{\mbox{\Large{\textit{a}}}_k}{\tau^{k+3}}
a^2 \left( 1 + \sum_{n=1}^\infty T_n^{(k)} a^n \right)
\end{equation}
where $a = a\left(\ln\frac{1}{\sqrt{\tau}\Lambda}\right)$ and $T_{n,0}^{(k)} \equiv T_n^{(k)}$. In eq. (53) there is no dependence on $\mu$.

As in the preceding sections, the $RS$ dependence of the expansion coefficients $T_n^{(k)}$ in eq. (53) can be found by using the equation
\begin{equation}\tag{54}
\left(\frac{\partial}{\partial c_i} + \sigma_i (a) \frac{\partial}{\partial a}\right) \mathcal{L}_k^{pert} = 0.
\end{equation}
With $\sigma_i(a)$ given by eq. (14), eq. (54) is satisfied order-by-order in $a$ provided
\begin{equation}\tag{55a}
\frac{\partial T_1^{(k)}}{\partial c_i} = 0 \Rightarrow T_1^{(k)} = \lambda_1^{(k)}
\end{equation}
\begin{equation}\tag{55b}
\frac{\partial T_2^{(k)}}{\partial c_i} + 2\delta_2^i = 0  \Rightarrow T_2^{(k)} = \lambda_2^{(k)} - 2c_2
\end{equation}
\begin{equation}\tag{55c}
\frac{\partial T_3^{(k)}}{\partial c_i} + 3 T_1^{(k)} \delta_2^i + \delta_3^i = 0  \Rightarrow T_3^{(k)} = \lambda_3^{(k)} - 3\lambda_1^{(k)} c_2 - c_3
\end{equation}
\begin{align}\tag{55d}
\frac{\partial T_4^{(k)}}{\partial c_i} &+ \left( 4 T_2^{(k)} + \frac{2}{3} c_2 \right)  \delta_2^i + \left( \frac{3}{2} T_1^{(k)} - \frac{c}{3} \right)\delta_3^i + \frac{2}{3} \delta_4^i = 0\\
& \Rightarrow T_4 = \lambda_4^{(k)}- 4 \lambda_2^{(k)} c_2 + \frac{11}{3} c_2^2 - \frac{3}{2} \lambda_1^{(k)} c_3 + \frac{cc_3}{3} - \frac{2}{3} c_4\nonumber
\end{align}
etc.\\
where $\lambda_i^{(k)}$ is a constant of integration and consequently is a $RS$ invariant that can be determined once $T_i^{(k)}$ and $c_i$ have been evaluated in some mass independent $RS$ such as $MS$.

Again, in the 't Hooft $RS$, $c_i = 0 \; (i \geq 2)$ and so $T_n^{(k)} = \lambda_n^{(k)}$.  In another $RS$, one can have $T_i = 0 (i \geq 2)$ so that by eq. (55)
\begin{equation}\tag{56a}
c_2 = \frac{1}{2} \lambda_2^{(k)}
\end{equation}
\begin{equation}\tag{56b}
c_3 = \lambda_3^{(k)} - 3\lambda^{(k)}_1 c_2
\end{equation}
\begin{equation}\tag{56c}
c_4 = \frac{3}{2} \left( \lambda_4^{(k)} - 4 \lambda_2^{(k)} c_2 + \frac{11}{3} c_2^2 - \frac{3}{2} \lambda_1^{(k)} c_3 + \frac{cc_3}{3}\right) 
\end{equation}
etc.\\
and $\mathcal{L}_k^{pert}(\tau)$ reduces to just two terms 
\begin{equation}\tag{57}
\mathcal{L}_k^{pert} = \frac{\mbox{\Large{\textit{a}}}_k}{\tau^{k+3}} a^2 \left( 1 + \lambda_1^{(k)} a\right).
\end{equation}

Using the approach taken in [18], we estimate the scalar glueball mass bound $\frac{\mathcal{L}_1}{\mathcal{L}_0}$ in the MS scheme for the 4-loop perturbative calculation with $\xi=\mu\sqrt{\tau} =0.8$ and 1.2, along with the $N^3LL$ RG summation and the scale independent result provided in eq. (53). The results are shown in Fig.2, where we find that remarkably the scale independent scalar glueball mass bound in the MS scheme is bounded within the $\mu$ dependent perturbative and RG summed results. In Figure 3, we depict the scale independent results for the mass bound for the MS, scheme 1 and scheme 2 as a function of $\tau$, where we find that the latter two schemes quite close to each other and all three schemes meeting at $\tau =$ 1 $GeV^{-2}$ which corresponds to a scalar glueball mass bound of $\frac{\mathcal{L}_1}{\mathcal{L}_0}\leq$ 1.2 $GeV$. 

\section{Discussion}

We have shown, by following the approach used in refs. [8-10], that if all logarithmic contributions to radiative corrections to QCD processes can be summed, then the explicit and implicit dependence on the renormalization scale parameter $\mu$ cancels. Furthermore, the $RS$ dependence can be analyzed and a $RS$ can be chosen so that either the $RG$ functions receive no contributions beyond two loop order, or the perturbative series in powers of the coupling terminates after a fixed number of terms.

Summing these logarithms is only possible if there is a sufficiently simple ansatz for the physical process being considered; it must be a power series in both the couplant and powers of the logarithm of $\mu$, with the power of the logarithm not exceeding that of the couplant (as in eqs. (1, 28, 47)).  This seems to require that we use a mass independent $RS$ such as $MS$.  An extension of the techniques that we have used so as to accommodate other renormalization schemes, especially when massive fields are  being considered, is currently being examined.  This would make it possible to consider electroweak processes, such as the semi-leptonic decay of the $b$ quark [10].

It would be most interesting to devise a way of summing the contribution of higher order effects in order to eliminate the explicit and implicit dependence on physical quantities on the RS dependent coefficients $c_i$. We are only able to arrange for the cancellation between the implicit and explicit dependence on $\mu$ because we know that the implicit dependence on $\mu$ takes a form such as appears in eqs. (1, 28, 47) on account of the way in which the appropriate Feynman diagrams contribute. There is apparently no analogous way of handling any of the $c_i$. 

We note that there are other approaches that have appeared in the literature in which the problem of reducing the $\mu$ dependence of results obtained using perturbative QCD has been addressed.  In particular, the ``principle of minimum sensitivity'' (PMS) invokes the principle (of which there is no proof) that the ``optimal'' value of unphysical parameters are those that minimize changes in the quantity being computed at some fixed order in perturbation theory when these parameters are varied. The original version of PMS [6] has been refined [20, 21] to accommodate higher order perturbative calculations. It would be of interest to see if the optimal values of $\mu$ and $c_i$ obtained by applying PMS at finite fixed order of perturbation theory lead to results compatible with the approach used in this paper in which an all-order summation has been used. One might also consider application of the PMS procedure to optimize the value of the $c_i$ after one has computed the RG summed result. 

In a second approach, which uses the ``principle of maximum conformality'' (PMC), dependence on the parameters $c_i(i \geq 2)$ is absorbed into the mass scale parameter $\mu$ at each order of perturbation theory [22-23]. This dependence of $c_i$ at each order of perturbation theory in PMC is found by direct ad hoc inspection of the computed value of the expansion coefficients that arise in the perturbative expansion of a physical quantity, rather than by using equations like eqs. (15, 35, 54). The dependence on $c_i$ is then absorbed into a mass scale parameter $\mu_n$, a different scale at each order of perturbation theory. This procedure does not make use of the RG summation employed in res. [8, 9] and in this paper to eliminate all dependence on the mass scale $\mu$ that arises due to renormalization; it is a procedure that is applied to a fixed order of perturbation theory and hence retains residual dependence on $\mu$. In ref. [27] it is demonstrated how one could use RG summation to reduce a perturbative expansion of a physical quantity to a power series in which the coefficients are RS independent and the running couplings are associated with same fixed scale (such as MS) but evaluated at each order of perturbation theory at a mass scale that absorbs all dependence on the RS parameters $c_i$ occurring at that order of perturbation theory. Results obtained using PMS and PMC have been compared for various processes in ref. [21].

The question of the infrared (IR) limit of results obtained using perturbative QCD can also be addressed when using the renormalization group summation employed in this paper.  One generally is concerned with the IR behaviour of $a(\mu)$ with $a(\mu)$ governed by eq. (2).  However, it is apparent that at least in perturbation theory the IR limit of $a(\mu)$ is contingent upon the RS used; the values of $c_i$ affect this limit if $b < 0$, at least when considering only a finite number of terms in the series expansion for $\beta(a)$.  A discussion of this IR behaviour in the scheme in which $c_i = 0$ (ie, $a(\mu)$ is the `t Hooft coupling of eq. (A.5) [15]) appears in ref. [24, 25]; the IR behaviour when the four loop contributions to $\beta(a)$ in the $\overline{MS}$ scheme is used appears in ref. [26].
 
It should be kept in mind that when examining the IR behaviour of a physical quantity computed using perturbation theory (such as $\mathcal{F}$ in eq. (1)), one should consider not just how $a(\mu)$ behaves in this limit, but also the convergence behaviour of the infinite series in powers of $a(\mu)$ that arises in perturbation theory.  Normally, the contribution of ``renormalons'' [15] is felt to result in such infinite series being at best asymptotic.  In ref. [27] it is proposed that, in order to circumvent the need to consider the convergence behaviour of infinite series, it is appropriate to simply examine the IR behaviour of such finite series as appear in eqs. (41) and (57).  In these cases the RG function $\beta(a)$ has the expansion coefficient $c_i$ fixed in terms of RS invariants ($\tau_i$ in the case of eq. (41) and $\lambda_i^{(k)}$ in the case of eq. (57)) and so this IR behaviour will depend on the process being considered.  In addition, if eq. (A.2c) is used to determine the value of $a$ in some RS in which the perturbative series does not terminate, it is likely that this value of $a$ is no longer a fixed point for the IR limit in this RS.  In ref. [27] this approach to examining the IR behaviour of the cross section for $e^+e^- \rightarrow$ (hadrons) results in a well defined IR limit for the cross section when there are $n_F = 3$ active flavours, even though the coupling $a$ diverges in the IR limit when it is computed using $\overline{M}S$ to four loop order.

\section*{Acknowledgements}
The suggestions of T.N. Sherry were most helpful, as was a comment by Roger Macleod.

\newpage

\begin{figure}
	\label{fig1}
\centerline{\includegraphics[height=14cm]{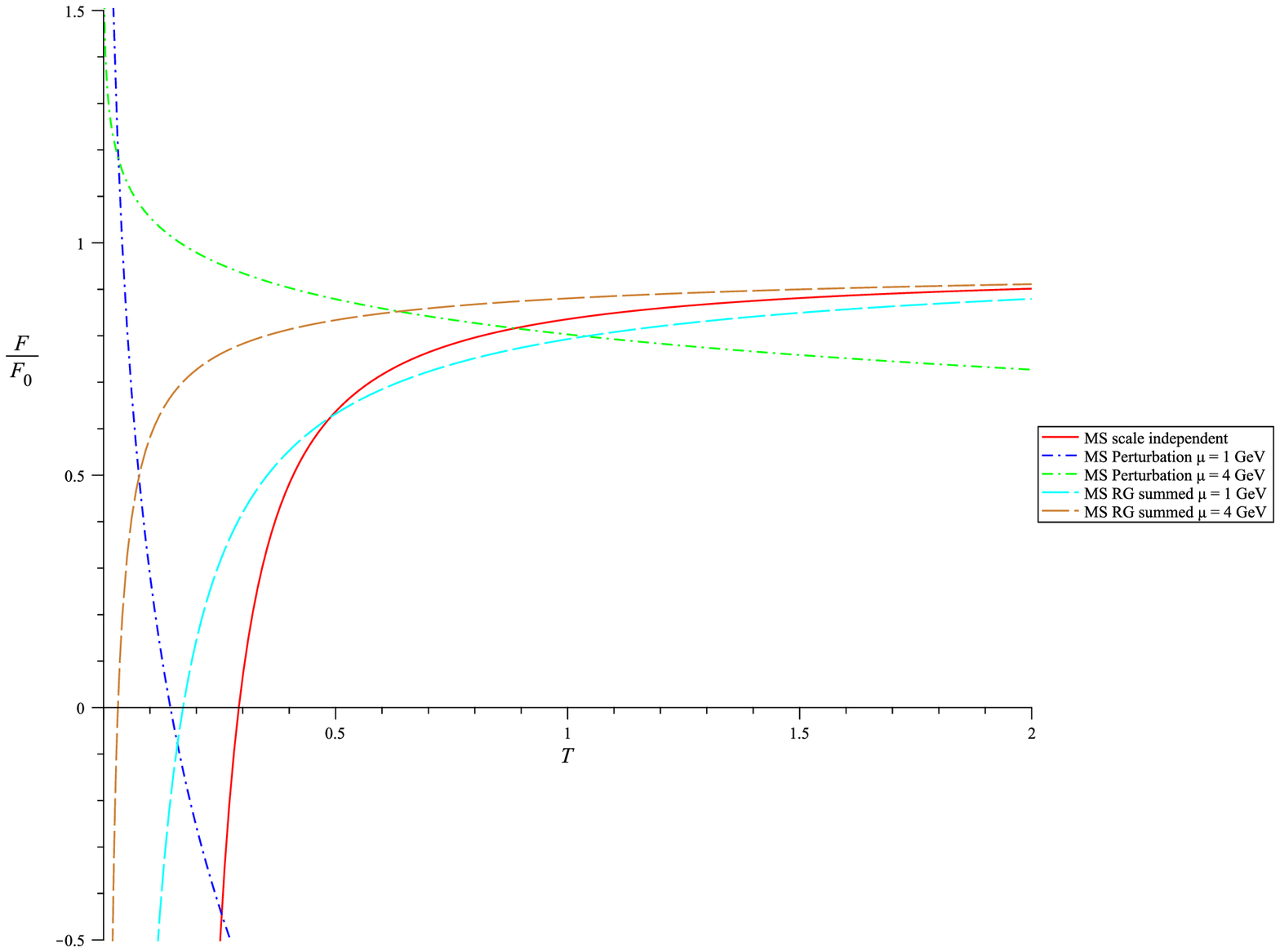}}
\caption{$\mathcal{F}/\mathcal{F}_0 $ as a function of temperature $T$ with perturbative and RG summed results}
\end{figure}

\begin{figure}
	\label{fig2}
\centerline{\includegraphics[height=14cm]{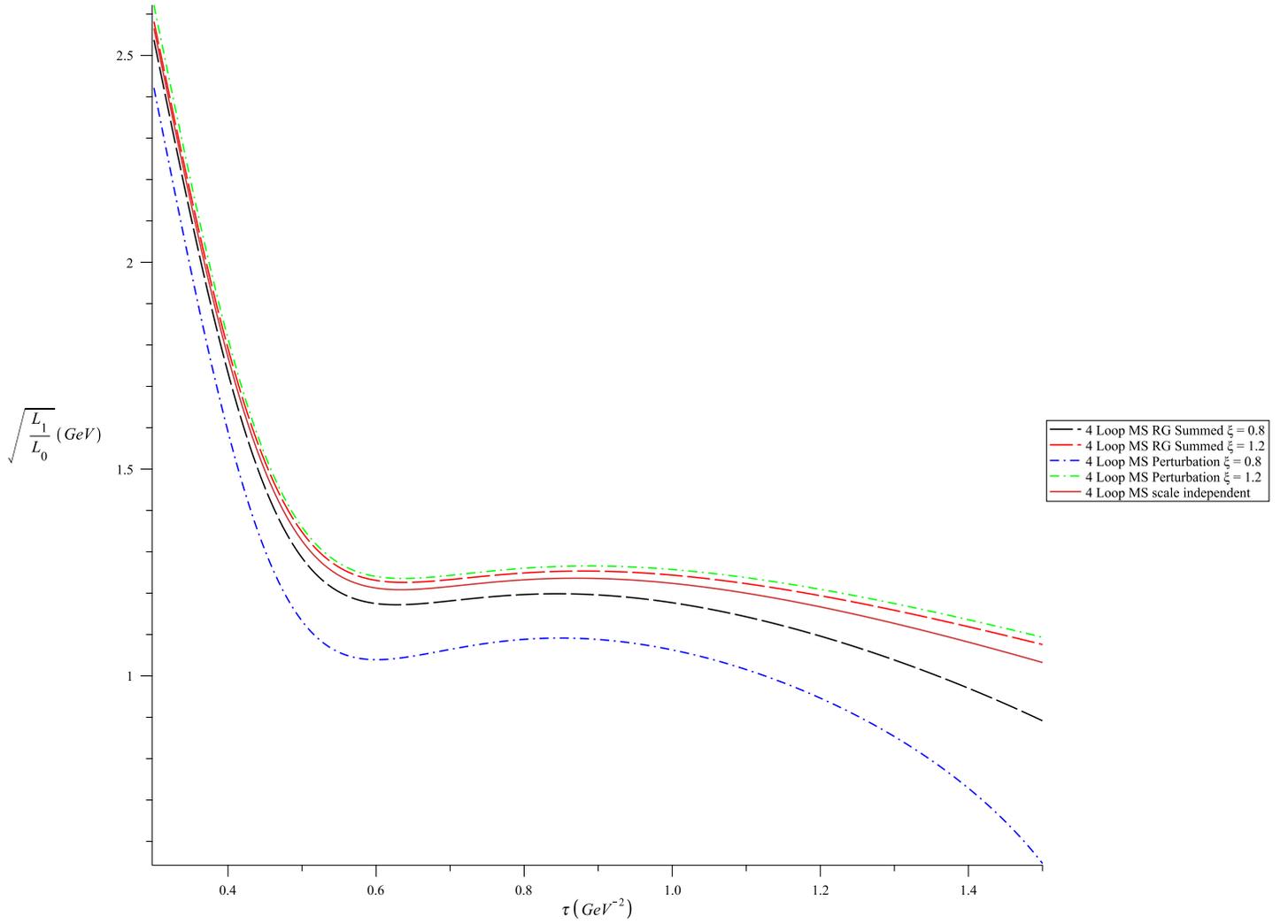}}
\caption{The $\mu$ dependence of the scalar glueball mass bound in MS scheme with truncated, RG summed and scale independent forms with respect to $\tau$($GeV^{-2}$) using values of $\xi= 0.8$ and $1.2$ respectively}
\end{figure}

\begin{figure}
	\label{fig3}
\centerline{\includegraphics[height=14cm]{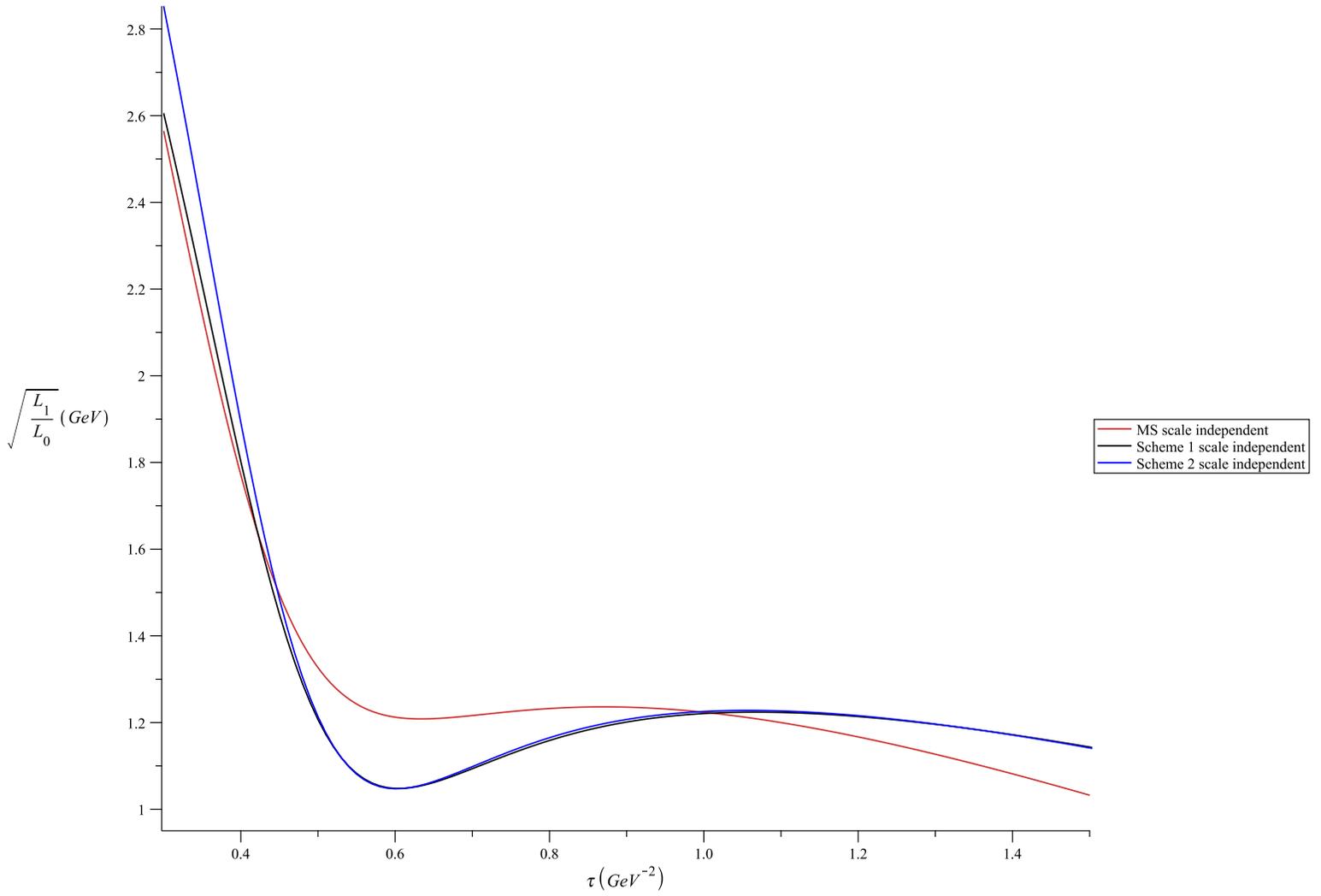}}
\caption{Scale independent scalar glueball mass bound as a function of $\tau$ in MS scheme, Scheme 1 and Scheme 2}
\end{figure}

\end{document}